\documentclass[10pt, conference, letterpaper]{IEEEtran}
\IEEEoverridecommandlockouts
\usepackage{cite}
\usepackage{amsmath,amssymb,amsfonts}
\usepackage{algorithmic}
\usepackage{floatrow}
\usepackage{graphicx}
\usepackage{textcomp}
\usepackage{xcolor}
\usepackage{multirow}
\def\BibTeX{{\rm B\kern-.05em{\sc i\kern-.025em b}\kern-.08em
    T\kern-.1667em\lower.7ex\hbox{E}\kern-.125emX}}
\usepackage{subfigure}  
\usepackage{rotating}
\usepackage{comment}

\begin{document}

\title{
	To Collaborate or Not in Distributed Statistical Estimation with Resource Constraints?
	\thanks{\fontsize{6.8}{7.8}\selectfont This research was sponsored by the U.S. Army Research Laboratory and the U.K. Ministry of Defence under Agreement Number W911NF-16-3-0001. The views and conclusions contained in this document are those of the authors and should not be interpreted as representing the official policies, either expressed or implied, of the U.S. Army Research Laboratory, the U.S. Government, the U.K. Ministry of Defence or the U.K. Government. The U.S. and U.K. Governments are authorized to reproduce and distribute reprints for Government purposes notwithstanding any copyright notation hereon. This project was partially sponsored by CAPES, CNPq and FAPERJ, through grants  E-26/203.215/2017 and E-26/211.144/2019.}
}
\author{
	\IEEEauthorblockN{Yu-Zhen Janice Chen\IEEEauthorrefmark{2}, Daniel S. Menasch\'e\IEEEauthorrefmark{1}, Don Towsley\IEEEauthorrefmark{2}}
	\IEEEauthorblockA{\IEEEauthorrefmark{2}College of Information and Computer Sciences, UMass, Amherst, Email: \{yuzhenchen, towsley\}@cs.umass.edu}
	\IEEEauthorblockA{\IEEEauthorrefmark{1}Department of Computer Science, UFRJ, Rio de Janeiro, Brazil, Email: sadoc@dcc.ufrj.br}
	\vspace{-18pt}}
\maketitle
\newtheorem{assumption}{\textbf{Assumption}}

\begin{abstract}
	We study how the amount of correlation between observations collected by distinct sensors/learners affects data collection and collaboration strategies by analyzing Fisher information and the Cramer-Rao bound.
	In particular, we consider a simple setting wherein two sensors sample from a bivariate Gaussian distribution, which already motivates the adoption of various strategies, depending on the correlation between the two variables and resource constraints.  
	We identify two particular scenarios: (1) where the knowledge of the correlation between samples cannot be leveraged for collaborative estimation purposes and (2) where the optimal data collection strategy involves investing scarce resources to collaboratively sample and transfer information that is not of immediate interest and whose statistics are already known, with the sole goal of increasing the confidence on an estimate of the parameter of interest. We discuss two applications, IoT DDoS attack detection and distributed estimation in wireless sensor networks, that may benefit from our results. 
\end{abstract}



\floatsetup[table]{capposition=top}
\newcommand{\STAB}[1]{\begin{tabular}{@{}c@{}}#1\end{tabular}}

\section{Introduction}\label{sec:introduction}
We consider a scenario where parameters of a multivariate distribution are learned from data samples collected from multiple sensors. These samples can come from sensors with different modalities or geographically separated sensors having the same modality. The samples can be communicated either to a central site with a single learner or between geographically distributed sites, each with its own learner. When there are constraints on the amount of information that can be passed around, there are tradeoffs between the amount of data and the type of estimation that should be performed by the learners.

We study two different settings, decentralized and centralized, that differ according to the location wherein learning occurs.
In the decentralized setting, we assume geographically dispersed learners each have access to one sensor, can collect data from their neighbors, and locally estimate parameter of interest.  
In the centralized setting, data is sent to a central data center where learning takes place. We consider the case of two sensors that take observations from a bivariate Gaussian distribution. In the distributed setting, each has a learner associated with it. In both cases there are constraints on the amount of data that can be communicated.  Although this is a simple setting, it motivates the adoption of a variety of different data collection strategies depending on the constraints. We frame the problem of designing data collecting and collaboration strategy as a problem of either maximizing Fisher information~\cite{cover1999elements} and/or minimizing the Cramer-Rao bound (CRB) to study how dependencies among data samples collected by distinct sensors affect learning strategies.

We pose the following question pertaining to collaborative parameter estimation: \emph{How should resource constraints determine what strategies learners (one in the centralized setting, two in the distributed setting)  should use?}
Our findings are summarized in Table~\ref{tab:estimator}, and indicate that the desired level of collaboration depends on three aspects, namely $(a)$ the available information, $(b)$  the considered setting (centralized or decentralized estimation) and $(c)$ the resource constraints.

\begin{table}[t]
	\vspace{-2pt}
	\centering
	\caption{Summary of results}
	\label{tab:estimator} 
	\scalebox{0.76}{	\begin{tabular}{l|p{0.34\linewidth}|p{0.34\linewidth}|p{0.34\linewidth}|}
			&\multicolumn{2}{|p{0.64\linewidth}|}{Learning a single unknown mean } & Learning two   \\ \cline{2-3}
			&Known correlation & Unknown  correlation   & unknown means\\\hline\hline
			{\multirow{1}{*}{{\rotatebox[origin=c]{90}{\parbox[c]{1.8cm}{\centering Without constraints}}}}}	& closed-form unbiased estimators, meet CRB   & closed-form unbiased estimators, do not meet CRB& closed-form unbiased estimators, meet CRB \\
			& \textbf{\emph{correlation always  helps}} &  \textbf{\emph{correlation may help}} &  \textbf{\emph{do not leverage correlation}} \\ \hline \hline
			{\multirow{4}{*}{{\rotatebox[origin=c]{90}{\parbox[c]{3cm}{\centering With constraints}}}}} &  \multicolumn{3}{l|}{Decentralized Learning Setting}\\ \cline{2-4}
			& closed-form  estimators, meet CRB  & numerical optimization & closed-form unbiased estimators, meet CRB \\\cline{2-4}
			&\multicolumn{3}{l|}{Centralized Learning Setting}\\ \cline{2-4}
			&numerical optimization &numerical optimization &closed-form unbiased estimators, meet CRB \\ \hline
	\end{tabular}	}  
	\vspace{-12pt}
\end{table}

Our contributions are threefold:\\
\emph{\textbf{Problem formulation.}} 
	We formulate a model that allows us to answer the question of what data is important
	which corresponds to minimizing a bound on the variance of the estimates, under constraints on the resource used for making observations, transmitting and receiving the samples.    
	In particular, we consider a bivariate Gaussian model, which contains five parameters, where two are the means. 
	We focus on the tasks of learning one or two means, under the assumption that some of the other parameters are known, and refer to the latter as side information.
	
\emph{\textbf{Extensive analysis.}} 
	For the unconstrained setup, we analytically show that, for estimating a single unknown mean, it is always worth leveraging available side information for the purposes of reducing estimation variance, as far as the correlation among samples is known; and when correlation is not known, we present conditions under which the use of side information is beneficial. When learning two unknown means, simple marginal estimators suffice, regardless of whether correlation is known. 
	Accounting for resource constraints, we numerically evaluate multiple data collecting strategies. 
	Our evaluation shows a number of interesting results, including: $(a)$ after accounting for constraints it may be beneficial to leverage side information for estimating the population means in all the considered scenarios; $(b)$  in particular, we identify cases where one needs to use different estimators depending on the amount of correlation between the considered observations.
	
\emph{\textbf{Applications.}} 
	We apply our model and results to two applications, distributed estimation in wireless sensor networks and the detection of distributed denial of service attacks.

In the rest of the paper, 
Section~\ref{sec:preliminary} describes the considered scenarios and problem formulation;
Section~\ref{sec:analysis} presents the problem solution in unconstrained and constrained scenarios;
Sections~\ref{sec:application} and \ref{sec:related-work} discuss potential applications and related work; and Section~\ref{sec:conclusion} concludes.


\floatsetup[table]{capposition=top}

\section{Problem Formulation}\label{sec:preliminary}
Consider two sensors, $S_x$ and $S_y$, which may or may not reside at the same location. We consider a time slotted system. At each time slot, sensors $S_x$ and $S_y$ can make independent observations on random variables $X$ and $Y$ respectively.  
Note that $X$ and $Y$ can represent different modalities or features, or have the same modality but be collected from  different geographic locations. 
Henceforth, we refer to a single observation from a single sensor as a \emph{marginal observation} or simply as an \emph{observation}, whereas a \emph{joint observation} comprises a pair of observations from the two sensors at the same time slot. A sample refers to either a marginal or a joint observation.  
In application, the two sensors may or may not belong to the same organization. In the case that the sensors belong to different organizations, our results shed insight on whether it is worthwhile for them to collaborate on different learning tasks.

\emph{\textbf{Observation Model and Tasks.}}
We model observations as coming from a bivariate Gaussian distribution, which includes five parameters, i.e., means $\mu_x$, $\mu_y$, and variances $\sigma_x^{2}$, $\sigma_y^2$, for $X$, $Y$, respectively, together with the Pearson correlation coefficient between $X$ and $Y$, denoted by $\rho$.
Our focus is on learning one or both means, under the assumption that some of the other parameters are known; we refer to those other known parameters as side information.
We consider three tasks: 
\begin{itemize}
	\item Task 1 (\textbf{T1}) is to learn one unknown mean, $\mu_y$, when all other parameters, including $\mu_x$, are known; 
	\item  Task 2 (\textbf{T2}) is to learn one unknown mean, $\mu_y$, when all parameters other than the correlation, $\rho$, are known;
	\item  Task 3 (\textbf{T3}) is to learn two unknown means, $\mu_x$ and $\mu_y$, when all other parameters are known. 
\end{itemize}
Henceforth, we refer to \textbf{T1} and \textbf{T2} as tasks involving the learning of a single unknown, and to task \textbf{T3} as a task involving the learning of two unknowns.

\emph{\textbf{Decentralized and Centralized Learning Settings.}}
We consider two learning settings.
In the {\em decentralized setting}, learners are placed at different geographic locations. 
Associated with each sensor is a learner that has the capability of making marginal observations, processing data samples locally, and transmitting data samples to the other learner for collaborative estimation. 
In the {\em centralized  setting}, the sensors transmit their observations to a central {\em data center} with a learner that  is responsible for estimating the unknown parameters; there is no local computation happening at the sensor nodes.

There is a resource cost associated with taking an observation and with transmitting it.  
We assume that the cost to make an observation is one unit and that the cost of communication per observation (either to transmit or receive it) is $\alpha$.
Consider, for instance, sensor $S_x$ that locally collect samples from $X$. If all of its observations are paired with observations from  $S_y$ to produce joint observations,
in the decentralized setting, it will incur an average resource cost of $\alpha+1$ per sample.
If $S_x$ also transmits its observations to $S_y$, the cost increases to $2 \alpha+1$ per sample. 

At each time slot, 
let $p_x$ (resp., $p_y$) be the probability that \emph{only} sensor $S_x$ (resp., $S_y$) is active, and \emph{only} observes $X$ (resp., $Y$).
Let $p_{xy}$ be the probability that a joint observation is made, i.e., both sensors are active. Note that, by definition, $p_x + p_y + p_{xy} \leq 1$.
We associate a resource constraint with each sensor, denoted by $E_1$ and in the centralized setting, a resource constraint of $E_2$ at the data center.  
Consider \textbf{T1}, for example, in the decentralized setting. Sensor $S_y$'s resource constraint is set to $p_y + \alpha p_x + (\alpha+1) p_{xy}\!\leq\!E_1$, where $p_y + p_{xy}$ units of resource are spent on making observation and  $\alpha p_x + \alpha p_{xy}$ units on receiving observations from $S_x$.

\begin{table}[t]
	\vspace{-9pt}
	\scalebox{0.76}{	\begin{tabular}{|p{0.21\linewidth}|p{0.97\linewidth}|}
			\hline
			Variable & Description \\
			\hline
			$(x, y) \in \mathbb{R}^2$ & Realization of a joint observation   \\
			$\mu_x, \mu_y \in \mathbb{R}$ & Means of random variables $X$ and $Y$ \\ 
			$\sigma_x^2, \sigma_y^2  \in \mathbb{R}$ & Variances of  $X$ and $Y$ \\
			$\rho \in [-1, 1]$ & Correlation coefficient, $\rho \sigma_x \sigma_y = \sigma_{xy}$ \\
			$p_x \in [0, 1]$ & Probability of sampling only a marginal observation from  $X$ \\
			$p_y \in [0, 1]$ & Probability of sampling only a marginal observation from $Y$ \\
			$p_{xy} \in [0, 1]$ & Probability of sampling joint observation \\
			$K \in \mathbb{N}$ & Total number of time slots\\
			$\alpha \in \mathbb{R}$ & Ratio of resource costs for communication  \\
			& (sending or receiving) over making an observation\\
			$E_1 \in \mathbb{R}$ & Average resource budget (sensors)\\
			$E_2 \in \mathbb{R}$ & Average resource budget (data center)\\
			\hline
	\end{tabular} }
	\caption{Table of notation}%
	\label{tab:notations}
	\vspace{-11pt}
\end{table}

\begin{figure*}[t]
	\vspace{-5pt}
	\setlength{\subfigcapskip}{-2.8pt}
	\centering
	\subfigure[Critical Points]{
		\includegraphics[width=0.27\textwidth]{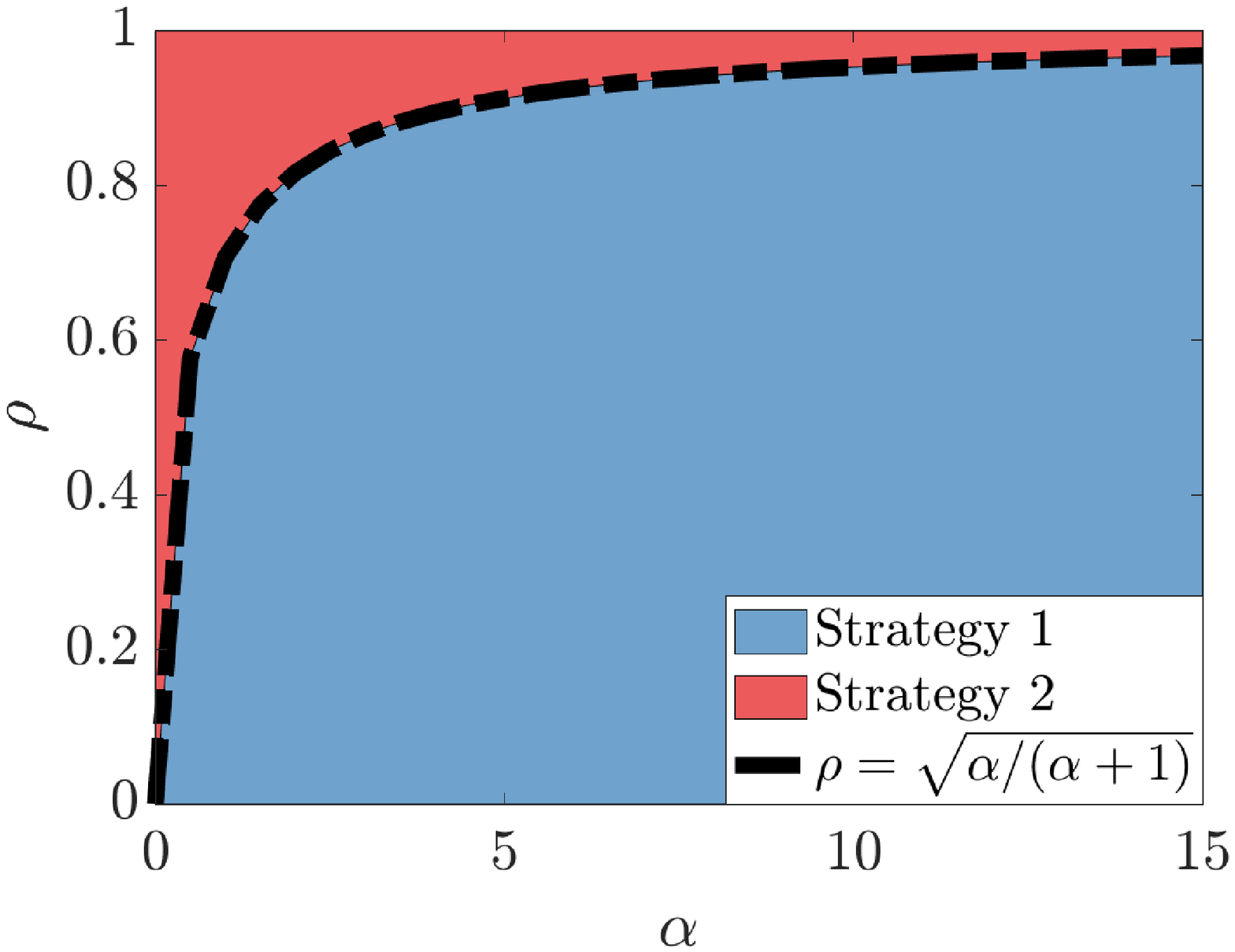}
		\label{fig:decentralized:known-rho-critical-points}
	}
	\subfigure[CRB: Constrained v.s. Unconstrained]{
		\includegraphics[width=0.27\textwidth]{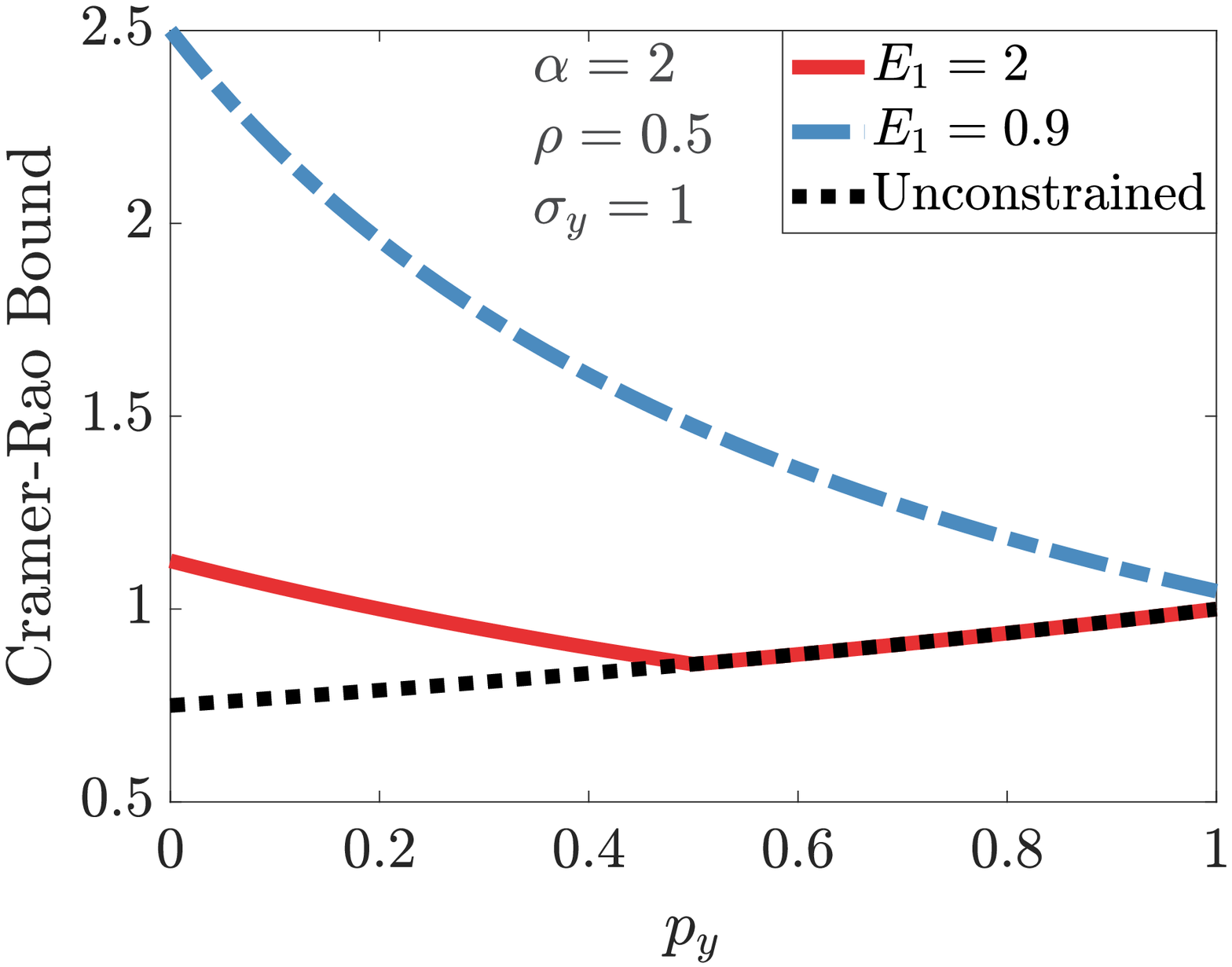}
		\label{fig:decentralized:known-rho-CRB}
	}
	\subfigure[Optimal Strategies]{
		\includegraphics[width=0.27\textwidth]{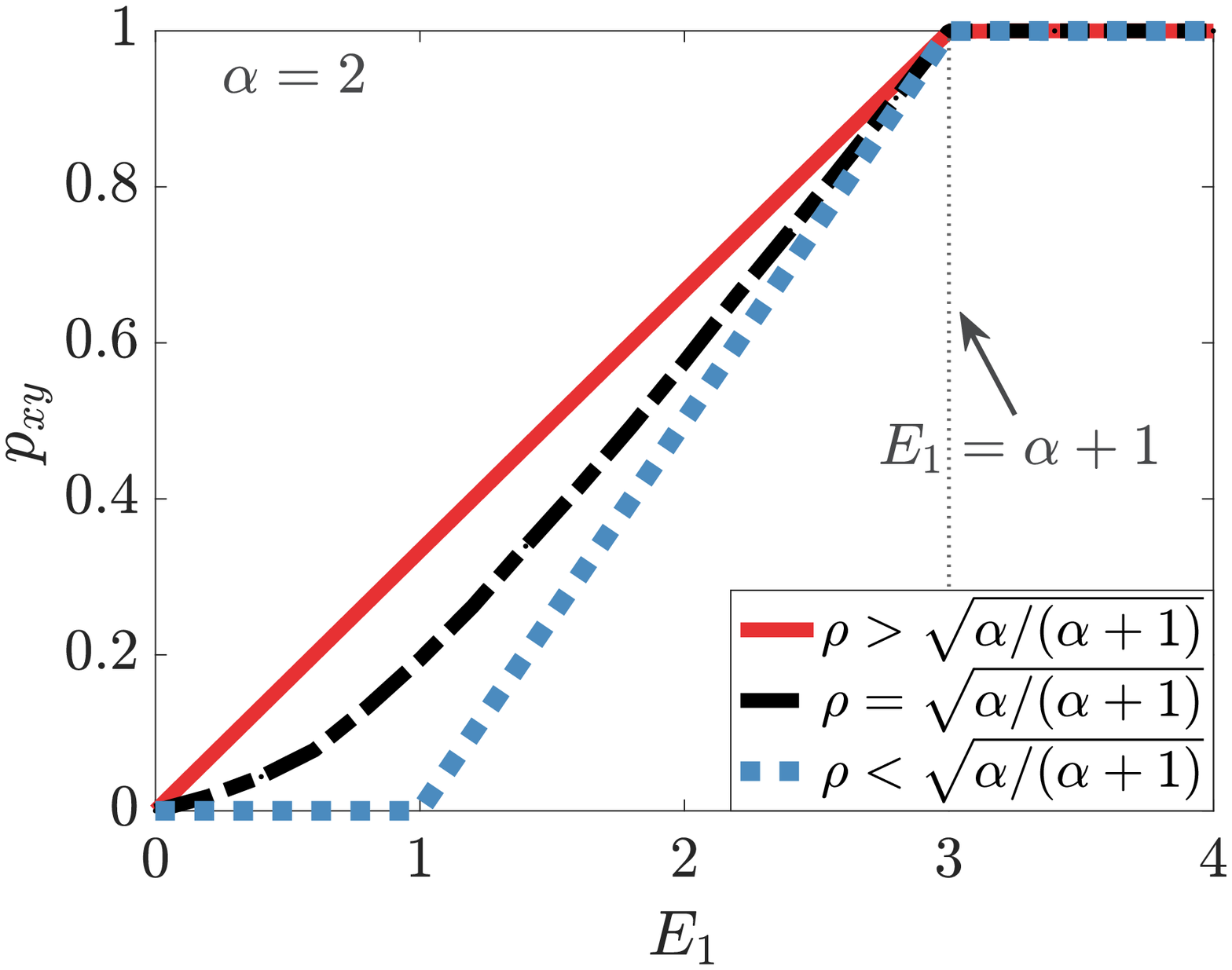}
		\label{fig:decentralized:known-rho-optimal-strategies}
	}
	\caption{Decentralized setting: \textbf{(T1)} learning a single unknown mean with correlation information subject to resource constraints. }
	\label{fig:decentralized:one-mean-unknown}
	\vspace{-7.8pt}
\end{figure*}
\section{Analysis}\label{sec:analysis}

\subsection{Decentralized Learning Setting}

\subsubsection{(\textbf{T1}) Leveraging Correlation Structure}\label{sec:decentralized:one-unknown:rho-known}
We begin by assuming that the correlation $\rho$ between $X$ and $Y$ is  known.
As we aim to estimate $\mu_y$, our objective is to maximize the Fisher information~\cite{cover1999elements} about $\mu_y$, given by $\mathcal{I}(\mu_y)$:
\begin{align}
&	\text{max } \,\,  p_y \mathcal{I}_Y(\mu_y) + p_{xy} \mathcal{I}_{(X,Y)}(\mu_y) = \frac{p_y }{\sigma_y^2} +\frac{ p_{xy} }{(1-\rho^2)\sigma_y^2}, \label{eq:decentalized:one-unknown:rho-known1}\\
&	\text{s.t. }\,\,  p_y +p_{xy} \leq 1. \label{eq:decentalized:one-unknown:rho-known2}
\end{align}
Note that maximizing the Fisher information corresponds to minimizing the Cramer-Rao bound, which bounds the variance of unbiased estimators.
For estimating $\mu_y$,  marginal observations from $X$, that are not paired up with observations of $Y$ to produce joint observations,  add no information,  $\mathcal{I}_X(\mu_y) = 0$. Therefore, $p_x = 0$.
It remains to determine whether to prioritize samples containing only information about $Y$ or joint samples.  In what follows, we show that the correlation coefficient plays an important role in that decision. Intuitively, the benefits from collecting joint samples ought to increase with correlation.  

\emph{\textbf{Prioritization.}} 
A joint sample, $(x, y)$, contains 
$\mathcal{I}_{(X,Y)}(\mu_y) =\frac{1}{(1-\rho^2)\sigma_y^2}$ 
Fisher information about parameter $\mu_y$. The resources used to collect $(x, y)$ can alternatively be used to collect $\alpha+1$ samples from  $Y$, which in total contain 
$\mathcal{I}_Y(\mu_y) = (\alpha+1)\frac{1}{\sigma_y^2}$ 
Fisher information about parameter $\mu_y$. Hence, when 
\begin{equation}
{1}/({(1-\rho^2)\sigma_y^2})>(\alpha+1)/{\sigma_y^2} \,\,\text{or}\,\, \rho^2>{\alpha}/({\alpha+1}), \label{eq:condikey}
\end{equation}
a joint observation, $(x, y)$, provides more information than $\alpha+1$ marginal observations from $S_y$.  The lower the resource cost, the lower the minimum value of $\rho$ that motivates collecting joint observations.   In particular, when there is no resource cost considered, 
$p_{xy}=1$ as $\mathcal{I}_{(X,Y)}(\mu_y) \geq \mathcal{I}_Y(\mu_y)$.   Figure~\ref{fig:decentralized:known-rho-critical-points} illustrates the regions where collecting marginal observations only from $Y$ (Strategy 1) or  joint observations (Strategy 2) should be prioritized. The dashed line separating the two regions corresponds to  $\rho=\sqrt{\alpha/(\alpha+1)}$.  

\emph{\textbf{Constrained maximization.}} 
Given the above prioritization scheme, we consider the constrained optimization problem~\eqref{eq:decentalized:one-unknown:rho-known1}-\eqref{eq:decentalized:one-unknown:rho-known2} with two additional resource constraints,
\[
p_z + (\alpha+1) p_{xy} \leq E_1, \quad (S_z\text{ resource constraint, } z\in\{x,y\})
\]
Figure \ref{fig:decentralized:known-rho-CRB} illustrates how these constraints affect the optimization problem and  the optimal data collection strategy. 
Let $\alpha=2$, $\sigma_y=1$ and $\rho=0.5$. 
When there are no resource constraints, i.e., $E_1 = \infty$ (black dotted line), $p_{xy} = (1-p_{y})$, and the Cramer-Rao bound (CRB)~\cite{cover1999elements} is minimized when $p_y=0$ and $ p_{xy}=1$; 
CRB is defined as the reciprocal of the Fisher information and serves as a lower bound on the variance of unbiased estimators.
When accounting for a resource budget of $E_1=2$ (red solid  line), the resource constraint of sensor $S_y$ becomes active in the region where $p_y \leq 0.5$; hence, for $0\leq p_y\leq0.5$, the value of $p_{xy}$ is determined by the $S_y$ resource constraint: $p_{xy} = (E_1-p_y)/(\alpha+1) \leq (1-p_{y})$. The CRB is minimized at $p_y=p_{xy}=0.5$. 
Under more stringent resource constraints (blue dashed line), data sharing becomes prohibitive and the CRB is minimized at $p_y=1$.

\emph{\textbf{Data collection strategies.}} The optimal data collection strategy is illustrated in Figure~\ref{fig:decentralized:known-rho-optimal-strategies}. 
When 
$\rho^2 < \alpha/(\alpha+1)$, 
the optimal strategy is to  make marginal observations on $Y$ (Strategy 1, denoted by the blue dotted line). Any residual resource budget is used to make joint observations. 
When 
$\rho^2 > \alpha/(\alpha+1)$, 
the entire resource budget should be used to collect joint observations  (Strategy 2, denoted by the red solid line). 
In summary, 
$$ p_{xy}=\begin{cases}
0, &\text{if }  \rho^2<\frac{\alpha}{\alpha+1}  \textrm{ and } E_1<1, \\
\frac{E_1-1}{\alpha}, & \text{if } \rho^2<\frac{\alpha}{\alpha+1} \textrm{ and }    1 \leq  E_1 <  \alpha +1, \\
\frac{E_1}{\alpha+1},  & \text{if }  \rho^2>\frac{\alpha}{\alpha+1} \textrm{ and }    E_1 < \alpha+1, \\
1,  & \text{if }    E_1  \ge \alpha+1.
\end{cases}$$ 
When 
$\rho^2=\alpha/(\alpha+1)$
the optimal value of $p_{xy}$ (denoted by the black dash-dot line in Figure \ref{fig:decentralized:known-rho-optimal-strategies}) are in between the values derived for the above two cases (i.e. between the red solid line and the blue dotted line). 

\emph{\textbf{Estimators.}} 
When Strategy 1 (prioritizing marginal observations) is adopted, we have $p_y >0$, $p_{xy} >0$. Following a methodology inspired by~\cite{wilks1932moments}, 
we obtain the estimator,
\begin{align}
\delta_1= \frac{(1-\rho^2)\bar{y}_1+\bar{y}-\beta\bar{x}}{2-\rho^2}, \label{eq:delta_1}
\end{align}
where $\bar{y}_1$ is the sample mean from marginal observations, and $\bar{y}$ and $\bar{x}$ are the sample means of joint observations.
The variance of $\delta_1$ per sample, $\text{Var}(\delta_1)$, is given by
\begin{align}
{\text{Var}(\delta_1)}\times{K}&=\frac{(1-\rho^2)\sigma_y^2}{(2-\rho^2)^2}\left(\frac{(1-\rho^2)}{p_y}+\frac{1}{p_{xy}}\right)(p_y+p_{xy}), \label{eq:var-delta_1}
\end{align}
where $K$ denotes the number of time slots, or, equivalently, the number of samples, as idle slots are assumed to correspond to empty samples for convenience.
The Cramer-Rao bound is given by 
\begin{align}
\mathcal{I}^{-1}(\mu_y) = \left((1-\rho^2)\sigma_y^2\right)/\left((1-\rho^2)p_y+p_{xy}\right). \label{eq:CRB-mu_y}
\end{align}
Comparing the above expressions,  $\delta_1$ achieves the CRB when $p_y=p_{xy}=0.5$; in that case,~\eqref{eq:var-delta_1} equals~\eqref{eq:CRB-mu_y}.  

When Strategy 2 (using all resources on joint observations) is adopted, $p_x=p_y=0$ and we apply the uniformly minimum-variance unbiased estimator (UMVUE) estimator,
\begin{equation}
\delta_2 = \bar{y} - \beta(\bar{x} -\mu_x),\,\,\, \beta = {\rho\sigma_y}/{\sigma_x},\label{eq:delta2}
\end{equation} 
to process the collected  joint observations. 

\begin{figure*}[t]
	\vspace{-5pt}
	\setlength{\subfigcapskip}{-2.8pt}
	\centering
	\subfigure[Decentralized Setting \textbf{T3}]{
		\includegraphics[width=0.27\textwidth]{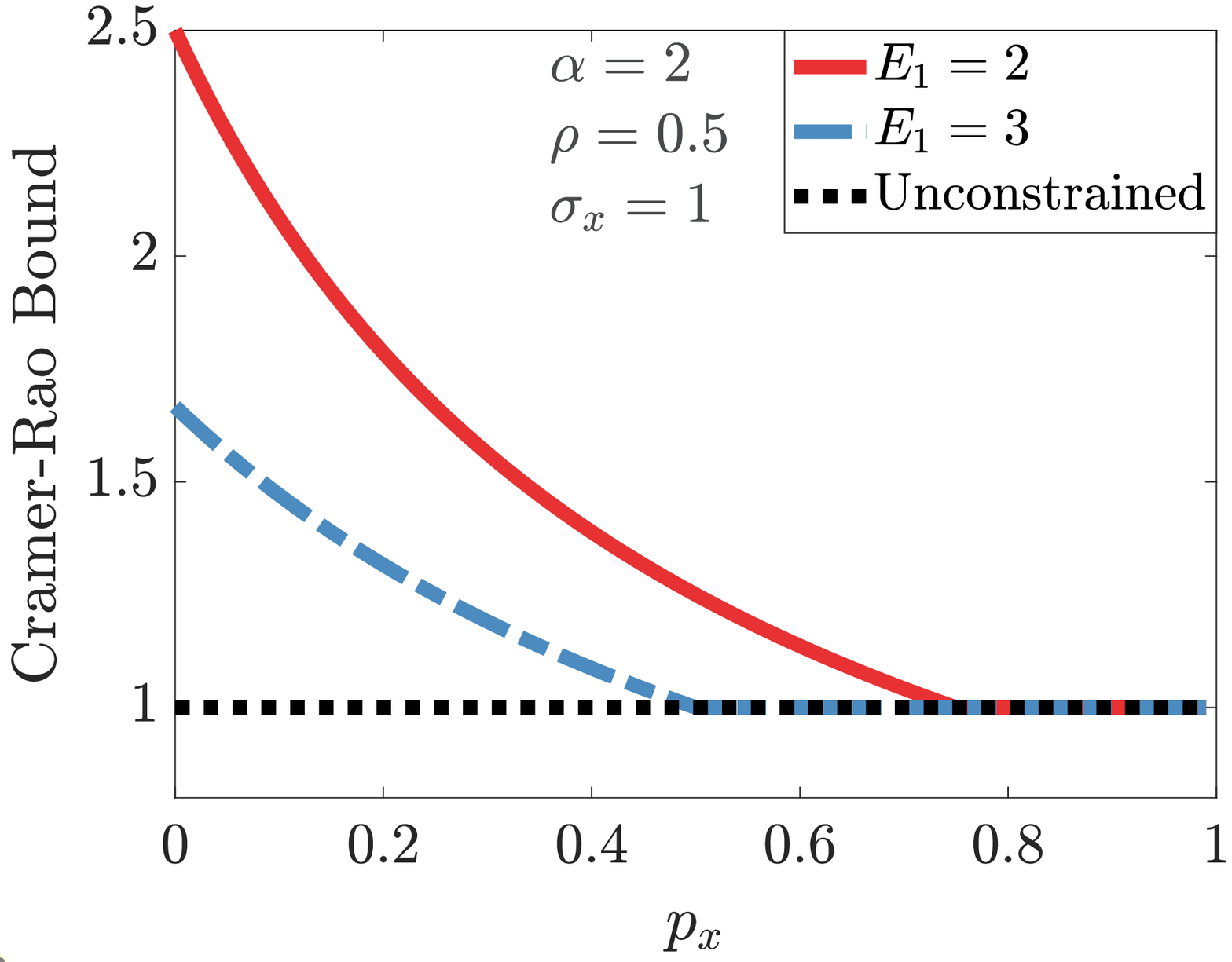} 
	}
	\subfigure[Centralized Setting \textbf{T1}]{
		\includegraphics[width=0.27\textwidth]{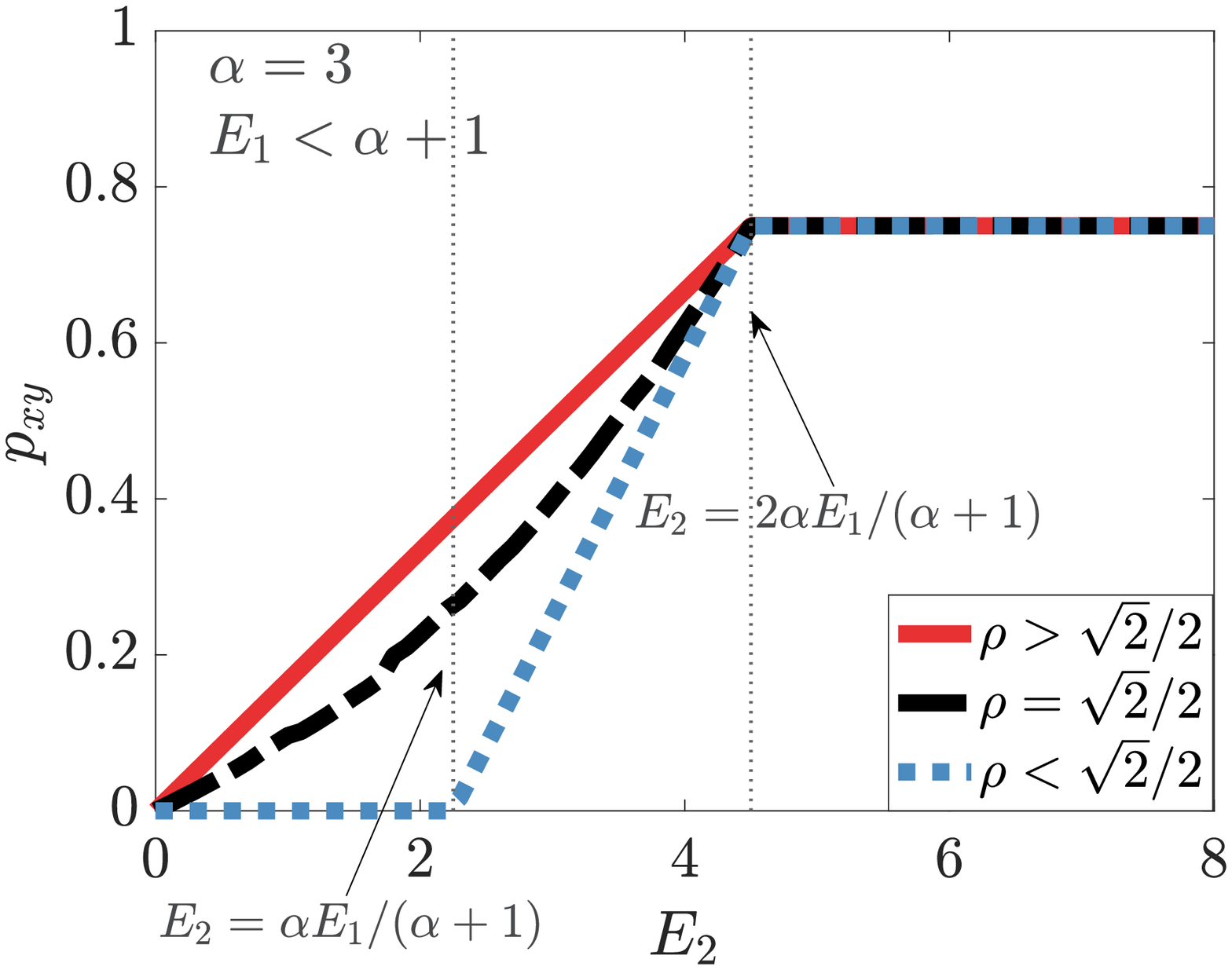} 
	}
	\subfigure[Centralized Setting \textbf{T1}]{
		\includegraphics[width=0.27\textwidth]{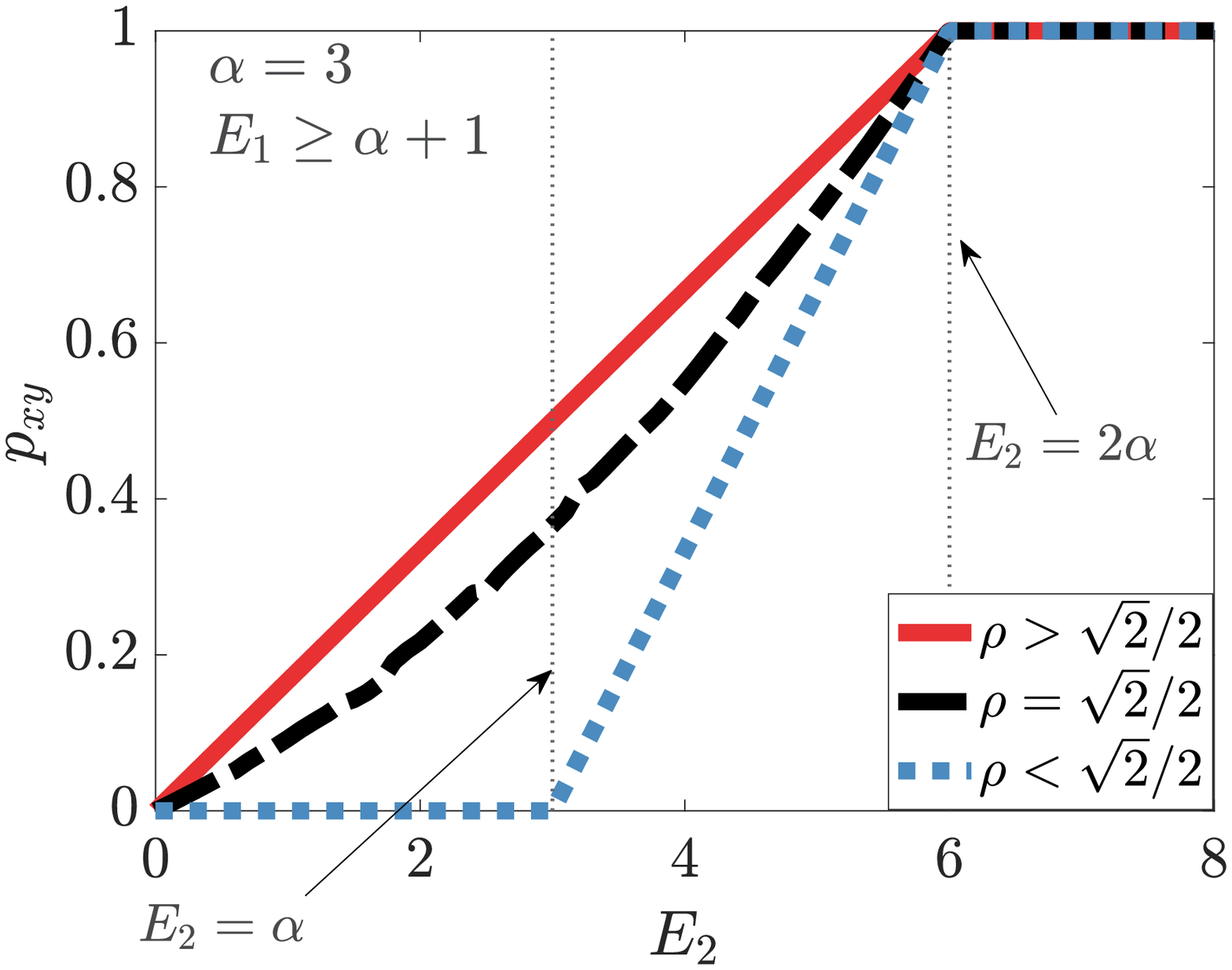}
	}
	\caption{ $(a)$ Decentralized setting, with two unknown means, where  $S_x$ aims at learning $\mu_x$, and $(b)$, $(c)$ centralized setting, with single unknown mean.}
	\label{fig:decentralized:two-mean-unknown-CRB}
	\vspace{-7.8pt}
\end{figure*}

\subsubsection{(\textbf{T2}) No Prior Correlation Information}\label{sec:decentalized:one-unknown:rho-unknown}
Consider the case where $\rho$ is unknown. As there are two unknown parameters, $\mu_y$ and $\rho$,   the Fisher information matrix (FIM) is given by:
\begin{align}
\mathcal{I}(\mu_y, \rho)=& p_y \mathcal{I}_Y\left(\mu_y, \rho\right) + p_{xy} \mathcal{I}_{(X,Y)}\left(\mu_y, \rho\right)\notag\\
=&
p_y\begin{bmatrix}
\frac{1}{\sigma_y^2} & 0\\
0 & 0
\end{bmatrix}+
p_{xy}
\begin{bmatrix}
\frac{1}{(1-\rho^2)\sigma_y^2} & 0\\
0 & \frac{1+\rho^2}{(1-\rho^2)^2}
\end{bmatrix}.
\end{align}
Our goal is to minimize the variance of the estimator of $\mu_y$, which is bounded by $\left[\mathcal{I}^{-1}\left(\mu_y, \rho\right)\right]_{11}$. As in this case the FIM is a diagonal matrix, the problem of minimizing
$ \left[\mathcal{I}^{-1}\left(\mu_y, \rho\right)\right]_{11} $ is equivalent to~\eqref{eq:decentalized:one-unknown:rho-known1}-\eqref{eq:decentalized:one-unknown:rho-known2}. 
For this case,~\cite{bishwal2008note} proposes an estimator which does not reach the CRB but is more efficient than the sample mean when $|\rho|  > 1/\sqrt{s-2}$. 
We refer the reader to~\cite{bishwal2008note} for details. 
When accounting for constraints, the prioritization and constrained maximization analysis introduced in the previous subsection still holds when $\rho$ is unknown.


\subsubsection{(\textbf{T3}) Learning Two Unknowns}\label{sec:constrained:decentralized:two-unknowns}
In this subsection, we study the scenario wherein two learners associated with $S_x$ and $S_y$ aim to learn $X$ and $Y$, respectively. 
All the other parameters are assumed to be known. The FIM is
\begin{align}
&\mathcal{I}(\mu_x, \mu_y)\notag\\
&=\!p_x \mathcal{I}_X(\mu_x, \mu_y) + p_y \mathcal{I}_Y (\mu_x, \mu_y) +p_{xy} \mathcal{I}_{(X,Y)}(\mu_x, \mu_y)\notag\\
&=\!p_x\! \begin{bmatrix}
\frac{1}{\sigma_x^2} & 0\\
0&0
\end{bmatrix}
\!+\! p_y\! 
\begin{bmatrix}
0&0\\
0&\frac{1}{\sigma_y^2}
\end{bmatrix}
\!+\! p_{xy}\!
\begin{bmatrix}
\frac{1}{(1-\rho^2)\sigma_x^2} & \frac{-\rho}{(1-\rho^2)\sigma_x\sigma_y}\\
\frac{-\rho}{(1-\rho^2)\sigma_x\sigma_y} & \frac{1}{(1-\rho^2)\sigma_y^2}
\end{bmatrix}.\notag 
\end{align}
\begin{align} \label{eq:inverse-two-mean}
&\mathcal{I}^{-1}(\mu_x, \mu_y) \\
&\quad= \frac{\sigma_x \sigma_y}{\Delta p_{xy}^3 (1-\rho^2)}
\begin{bmatrix}
\frac{({1-\rho^2})p_y+p_{xy}}{\sigma_y^2} & \frac{ \rho p_{xy} }{\sigma_x \sigma_y } \\
\frac{ \rho p_{xy} }{\sigma_x \sigma_y } & \frac{({1-\rho^2})p_x+p_{xy}}{\sigma_x^2}\notag
\end{bmatrix},\\
& \Delta = \frac{ 1+(p_x/p_{xy})+(p_y/p_{xy})+(p_x p_y/p^2_{xy}) (1-\rho^2)}{p_{xy} \sigma_x \sigma_y (1-\rho^2)}. \label{eq:deltaini}
\end{align}
In a decentralized learning setting, each learner aims to estimate its corresponding mean.  
Learner associated with $S_x$ aims to estimate the unknown mean $\mu_x$. Hence, its objective is to minimize the $\left[\mathcal{I}^{-1}(\mu_x, \mu_y)\right]_{11}$
\begin{align}
&\text{min }  \left[\mathcal{I}^{-1}(\mu_x, \mu_y)\right]_{11} \leq K \text{Var}(\hat\mu_x) \label{eq:unconstra21}\\
&\text{s.t. } p_x + p_y +p_{xy} \leq 1. \label{eq:unconstra22}
\end{align}
Similarly,  learner associated with $S_y$ aims at  minimizing the 2,2-entry
of the inverse of FIM, which corresponds to the above problem,  replacing 
$K\text{Var}(\hat\mu_x)$ by $K\text{Var}(\hat\mu_y)$.  

\emph{\textbf{Multidimensional Extension.}} 
Interestingly, it can be verified from the above expressions that  
the maximum likelihood estimators are the sample means in this scenario, which do not depend on the variances and covariances. 
Considering the probability density function of the $k$-variate Gaussian distribution, 
the above  result  also holds, 
i.e.,  
when all means are unknown, and the covariance matrix is known, there is no advantage to collaborate across sensors for estimation purposes. 
The maximum likelihood estimators are the sample means as shown in the following.
The probability density function of  $k$-variate normal distribution is 
$\mathcal{N}(\mathbf{x}_1,...\mathbf{x}_n|{\pmb\mu}, \mathbf{\Sigma})\equiv(2\pi)^{\frac{-k}{2}} \det (\mathbf{\Sigma})^{\frac{-1}{2}} \exp\left\{-\frac{1}{2}(\mathbf{x}-{\pmb\mu})^\top \mathbf{\Sigma}^{-1} (\mathbf{x}-{\pmb\mu})\right\}.$
Hence, the log-likelihood function is $\mathcal{L}({\pmb\mu}, \mathbf{\Sigma}|\mathbf{x}_1,...\mathbf{x}_n) = -\frac{nk}{2}\ln (2\pi)-\frac{n}{2} \ln (\det (\mathbf{\Sigma})) - \frac{1}{2} \sum_{i=1}^{n}(\mathbf{x}_i-{\pmb\mu})^\top \mathbf{\Sigma}^{-1} (\mathbf{x}_i-{\pmb\mu}).$ 
\begin{align*}
&\text{Let } \nabla_{\pmb\mu} \mathcal{L} ({\pmb\mu}, \mathbf{\Sigma}|\mathbf{x}_1,...\mathbf{x}_n) \\
&\!=\!\nabla_{\pmb\mu}\!\left[\!-\frac{nk}{2}\ln (2\pi)\!-\!\frac{n}{2} \ln (\det (\mathbf{\Sigma}))\!-\!\frac{1}{2} \sum_{i=1}^{n}\!(\mathbf{x}_i\!-\!{\pmb\mu})^\top \mathbf{\Sigma}^{-1} (\mathbf{x}_i\!-\!{\pmb\mu})\!\right]\!\\
&\!=\! - \sum_{i=1}^{n} \mathbf{\Sigma}^{-1} (\mathbf{x}_i\!-\!{\pmb\mu})\!=\! -\mathbf{\Sigma}^{-1} \sum_{i=1}^{n} (\mathbf{x}_i\!-\!{\pmb\mu})\!=0.
\end{align*}
The maximum likelihood estimator for ${\pmb\mu}$ is $\hat{\pmb\mu} = \frac{1}{n}\sum_{i=1}^{n} \mathbf{x}_i$.

Now, when taking resource budgets into account, the problem faced by sensor $S_x$ includes the following additional constraints,
\begin{align}
&	p_x + (2\alpha+1) p_{xy} \leq E_1, \quad (S_x\text{ resource constraint}) \label{eq:two-unknown:constraint1}\\
&	p_y + (2\alpha+1) p_{xy} \leq E_1, \quad  (S_y\text{ resource constraint}). \label{eq:two-unknown:constraint2}
\end{align}
Sensor $S_y$'s problem can be similarly defined.
Figure~\ref{fig:decentralized:two-mean-unknown-CRB}(a) shows how the CRB varies as a function $p_x$ depending on the constraints, where the learner associated with $S_x$ aims to learn $\mu_x$ when $\sigma_x=1$, $\rho=0.5$, $\alpha=2$.
Without resource constraints, any value of $p_x$, $ 0 \leq p_x \leq 1$, corresponds to an optimal solution; $p_x  + p_{xy} = 1$ and the optimal estimator is the sample mean, which attains the CRB.  When resource constraints are taken into account, the sampling of joint observations competes against the sampling of $X$, i.e., the resource cost of receiving observations from $S_y$ may preclude making additional observations on $X$. 
In that case,  $p_x$ must be set large enough so as to achieve the same efficiency as in the unconstrained scenario.  
For constrained scenario, one can apply the estimators derived in~\cite{wilks1932moments}, which meet the CRB.


\subsection{Centralized Learning Setting}

\subsubsection{(\textbf{T1}) Leveraging Correlation Structure}\label{sec:centralized:one-unknown:rho-known}
The problem of learning a single unknown when $\rho$ is known is posed in~\eqref{eq:decentalized:one-unknown:rho-known1}-\eqref{eq:decentalized:one-unknown:rho-known2}. 
In a centralized setting, sensors need to transmit every sample they obtain to the central data center (DC). The resource constraints are given by:
\[
(\alpha\!+\!1)(p_z\!+\!p_{xy}) \leq E_1, \quad (S_z\text{ resource constraint, }z\in\{x,y\})
\]
We also admit resource constraints related to the DC, which spends resource to receive the samples: 
\begin{align*}
& \alpha(p_x\!+\!p_y\!+\!2 p_{xy}) \leq E_2. &(\text{DC resource constraint})
\end{align*}
Focusing on the DC constraint, the decision to prioritize collection of marginal or joint observations is similar to that considered in Section~\ref{sec:decentralized:one-unknown:rho-known}.  
In particular, resources consumed to collect one $(x,y)$ sample are assumed to be twice the amount of resources needed to collect one sample from $Y$. 
Therefore, the DC problem can be tackled as a special case
of the problem considered in Section~\ref{sec:decentralized:one-unknown:rho-known}, with $\alpha=1$. 
The DC should  prioritize joint observations if $\rho > \sqrt{2}/2$,
noting that the condition follows  from~\eqref{eq:condikey} by letting $\alpha=1$.

Figure~\ref{fig:decentralized:two-mean-unknown-CRB}(b) illustrates the optimal data collection strategy as a function of DC resource budget, $E_2$, in a scenario where  the resource constraint of $S_y$ is active.  When $\rho<\sqrt{2}/2$, it is beneficial to prioritize the collection of marginal observations,  and only collect joint observations  when there are spare resources (as illustrated by the blue dotted lines). 
When $\rho > \sqrt{2}/2$, in contrast, it is beneficial to prioritize joint observations.
In Figure~\ref{fig:decentralized:two-mean-unknown-CRB}(c), the resource constraint on $S_y$ is inactive, i.e., the budget $E_1$ is increased. 
When $E_1$ and $E_2$ are large, Figure~\ref{fig:decentralized:two-mean-unknown-CRB}(c) shows that $p_{xy}=1$.
Figures~\ref{fig:decentralized:two-mean-unknown-CRB}(b) and~\ref{fig:decentralized:two-mean-unknown-CRB}(c) show that the maximum attainable value of $ p_{xy}$ increases as $E_1$ increases, and as far as  $E_1 \ge \alpha+1$ ($S_y$ constraint inactive), $p_{xy}$ eventually reaches 1 given large enough $E_2$. 

\subsubsection{(\textbf{T2}) No Prior Correlation Information}
When $\rho$ is not known in advance, our problem formulation is equivalent to~\eqref{eq:decentalized:one-unknown:rho-known1}-\eqref{eq:decentalized:one-unknown:rho-known2}, with the same constraints as considered in Section~\ref{sec:centralized:one-unknown:rho-known}.  In particular,  the analysis and the results presented above still hold. Nonetheless,  the  derivation and analysis of   estimators targeting the obtained bounds, without leveraging $\rho$, are left as subjects for future work.

\subsubsection{(\textbf{T3}) Learning Two Unknowns}
\begin{figure}[!th]
	\vspace{-6pt}
	\includegraphics[width=0.549\textwidth]{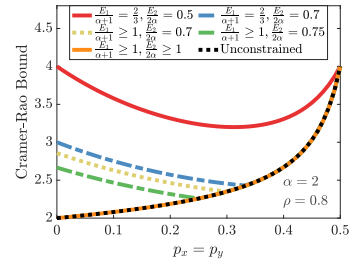}
	\caption{Centralized setting: \textbf{(T3)} Learning two unknown means.} 
	\label{fig:centralized:two-mean-unknown}
	\vspace{-2pt}
	\vskip -0.05 in
\end{figure}

Next, we consider the problem of learning two correlated unknowns accounting for the resource constraints introduced in Section~\ref{sec:centralized:one-unknown:rho-known}.  
Recall that without constraints the sample means were shown to achieve the CRB. By accounting for the constraints, in contrast, we obtain a much richer class of sampling strategies of interest.  Figure~\ref{fig:centralized:two-mean-unknown} illustrates this point through a simple example, with $\alpha=2$ and $\rho=0.8$. In the unconstrained case, the CRB is minimized when $p_x=p_y=0, p_{xy} = 1$.   
When resource constraints are considered, the data collection strategy that minimizes the CRB may involve setting $p_x>0$, $p_y>0$ as well as $p_{xy}>0$. Note that by definition $p_x + p_y + p_{xy} \leq 1$. The optimal value of $p_{x}$ varies as a function of the resource constraints, noting that stricter budgets translate into incentives favoring marginal observations.

\begin{figure*}[t]
	\vspace{-5pt}
	\setlength{\subfigcapskip}{-2.8pt}
	\centering
	\subfigure[Stringent Constraints]{
		\includegraphics[width=0.27\textwidth]{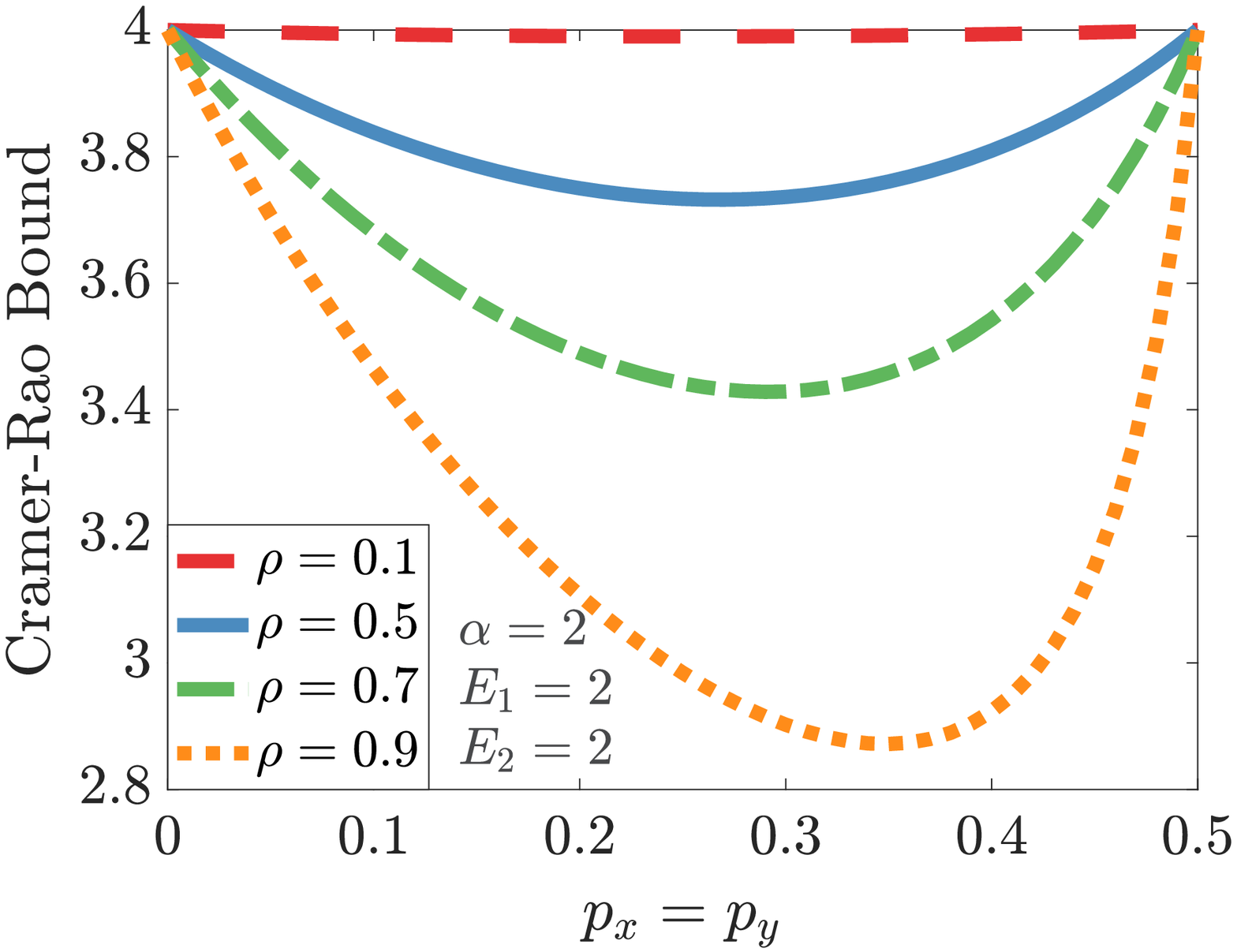}
	}
	\subfigure[Stringent Constraints - Effect of $\rho$]{
		\includegraphics[width=0.27\textwidth]{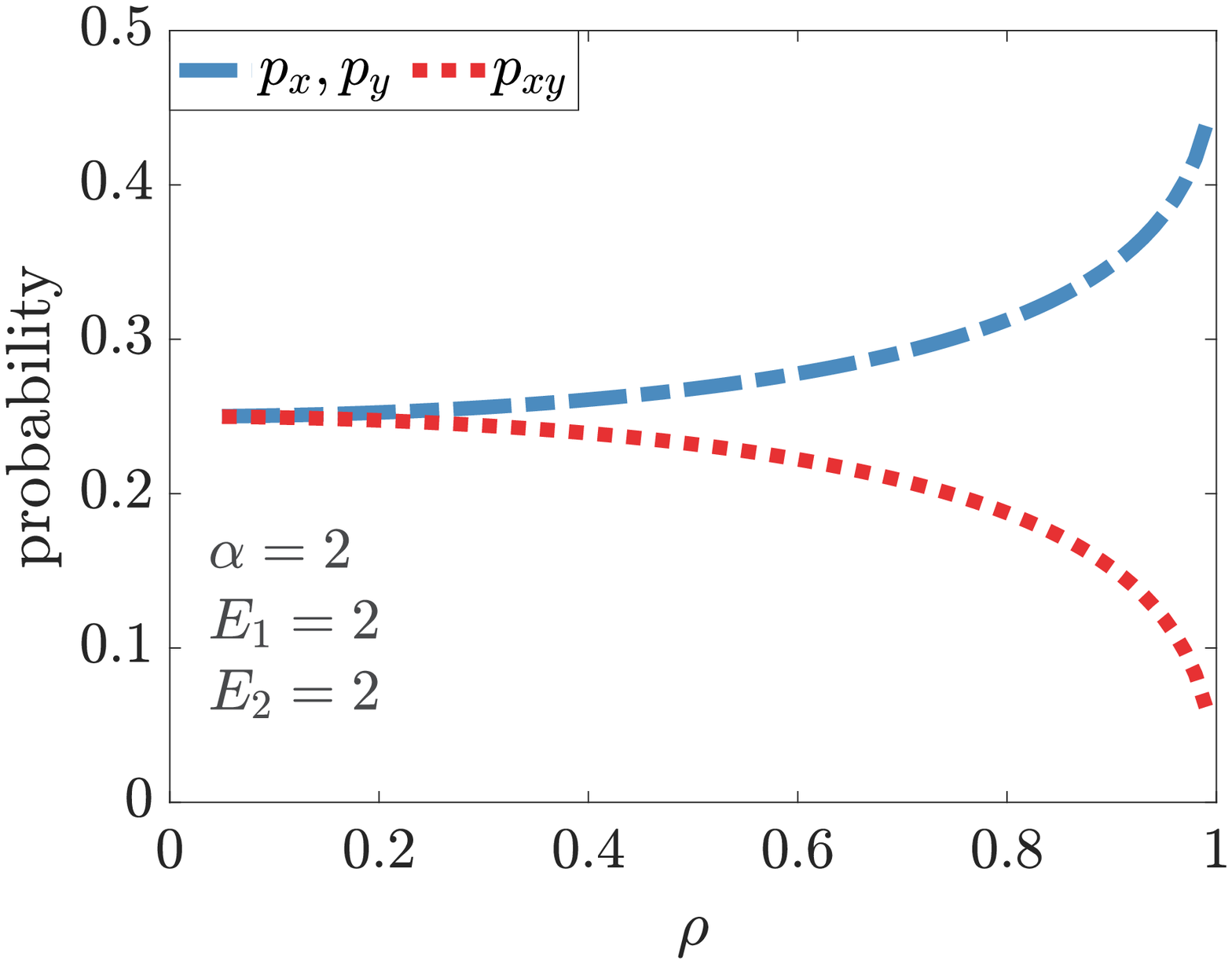}
	}
	\subfigure[Relaxing Constraints]{
		\includegraphics[width=0.27\textwidth]{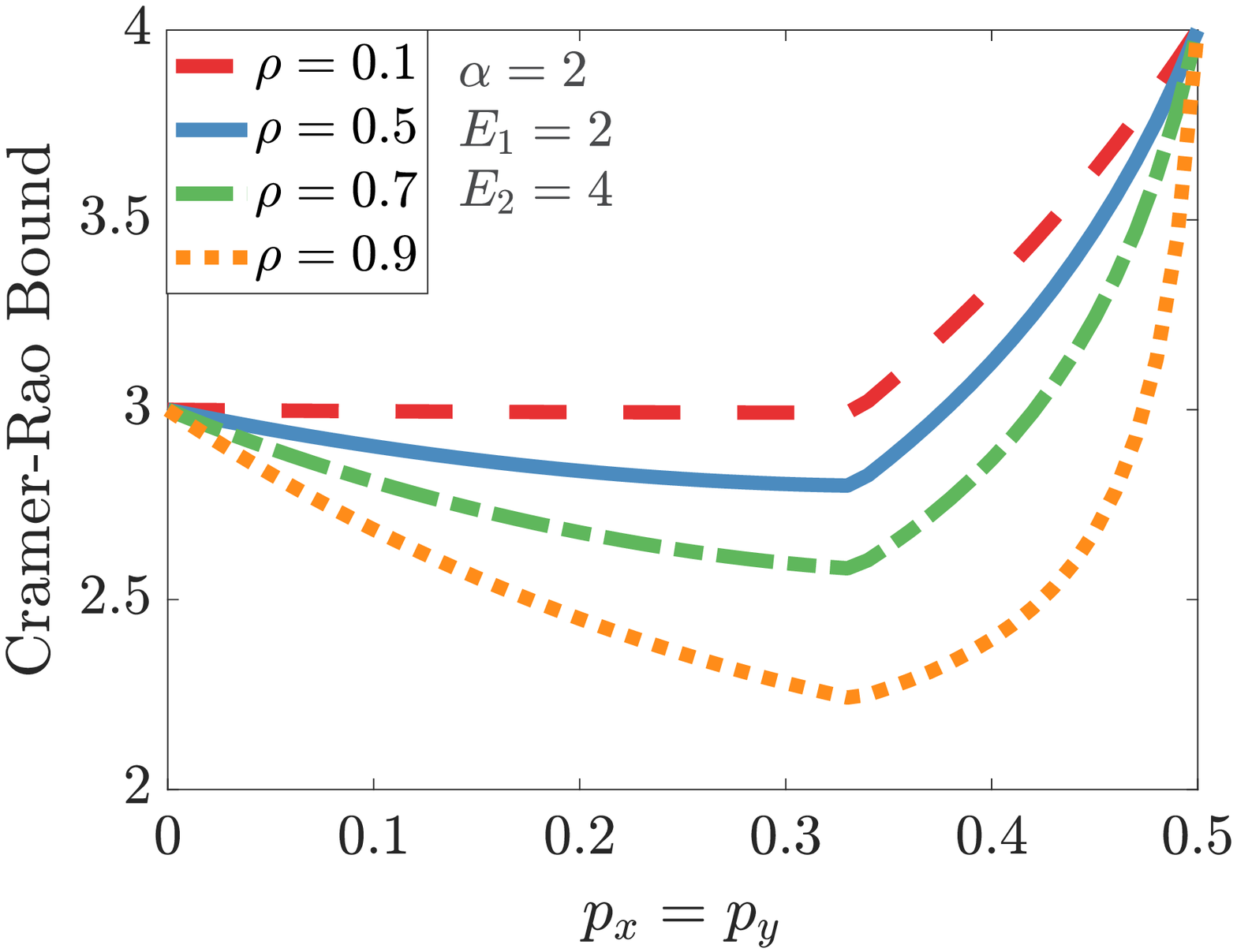}
	}
	\caption{Centralized setting: \textbf{(T3)} learning two unknown means, varying $\rho$.}  
	\label{fig:twocentral}
	\vspace{-7.8pt}
\end{figure*}

Figure~\ref{fig:twocentral}  indicates how the optimal policy varies as a function of $\rho$.  In Figure~\ref{fig:twocentral}(a) we let $\sigma_x=\sigma_y=1$, $E_1=E_2=2$ and $\alpha=2$.  
First, note that the sensitivity of the CRB with respect to $p_{xy}$ grows as a function of $\rho$.
Second, as $\rho$ grows the value of $p_x=p_y$ which minimizes the CRB increases, i.e.,    
the larger the correlation, the less often should   joint observations be sampled.
Figure~\ref{fig:twocentral}(b) shows the optimal strategy as a function of $\rho$, indicating that as $\rho$ approaches 1, marginal observations should be sampled more often. 
In Figure~\ref{fig:twocentral}(c) we relax the resource constraints, causing a reduction in the CRB specially for values of $p_x < 0.3$.  

\section{Applications}\label{sec:application}
Extensive sampling from the physical environment, e.g., air, water and surface sampling, and from virtual ecosystems, e.g., network traffic, and collaboratively learning a characteristic parameter about the environment is a landmark of modern computer and communication systems~\cite{romer2004design, akyildiz2002wireless}.
Its applications range from smart home monitoring and military coalition to environmental monitoring, where characteristic parameters may be  temperature, GPS-related signals and air pollution, respectively~\cite{zhao2003collaborative}.
In the following, we describe two applications where our model and analysis can be applied.

\emph{\textbf{IoT DDoS Attack Detection.}}
As the number of devices connecting to home networks, e.g., PCs, tablets, mobile devices and IoT devices like smart thermostats, keeps increasing in recent years, it has attracted the attention of malicious agents interested in compromising those devices and launching distributed denial of service (DDoS) attacks~\cite{marzano2018evolution}. Many Internet service providers have installed software at home routers that are used to periodically make a variety of observations such as numbers of packets and bytes uploaded and downloaded. These observations can be used to estimate their means and/or correlations. One can model this as a collection of multimodal sensors in a home router and/or a set of sensors of the same modality distributed across homes.  Data from these sensors are then collected at a data center subject to constraints on available bandwidth from home router to the data center.  Such a design has been used to develop detectors for DDoS attacks~\cite{mendoncca2019extremely}.


\emph{\textbf{Distributed Estimation in Wireless Sensor Network.}}  
In wireless sensor networks (WSNs), energy is typically the critical resource and communication usually dominates energy consumption of embedded networked systems whose components have limited on-board battery power~\cite{zhao2003collaborative}, posing the challenges of determining whether devices should collaborate or not, and setting the rate at which information must be transmitted through the network given the metrics to be estimated. 
Sensors in WSNs map naturally to sensors in our model.
If sensors all sense the same variable, there are usually certain spatial correlations between observations~\cite{vuran2006spatial}; depending on whether the controller of the sensors has sufficient prior knowledge about the correlation structure, learning tasks map to our \textbf{T1} or \textbf{T2}. If sensors sense different modalities~\cite{zhao2003collaborative}, our results about \textbf{T2} or \textbf{T3} may also apply.

\section{Related Work}\label{sec:related-work}

Our problem formulation is inspired by the literature on  inference of parameters of  the Gaussian distribution. In particular, some of the early results on the amount of information contained in a sample with missing data, derived by Wilks~\cite{wilks1932moments} and later on extended by Bishwal and Pena~\cite{bishwal2008note}, serve as foundations for our search for optimal data collecting strategies accounting for maximum likelihood estimators.  
Whereas~\cite{wilks1932moments, bishwal2008note} assume that one has no control over missing data, for designing optimal data collecting strategies, the goal is to determine which data must be ``missed", e.g., due to resource constraints.  By leveraging this observation,  we build on top of~\cite{wilks1932moments, bishwal2008note}, posing the design of collaborative estimation strategy as a constrained optimization problem and deriving properties of its solution.

Fisher information has been widely applied in the realm of computer networks, e.g., assessing the fundamental limits of flow size estimation~\cite{ribeiro2006fisher}, network tomography~\cite{he2015fisher} and sampling in sensor networks~\cite{song2009optimal}.
Our methodology and results differ from previous work in at least two aspects. 
One, we consider both centralized and decentralized scenarios and indicate the key role played by resource constraints in each scenario. 
Two, we assume that samples are collected from a bivariate Gaussian distribution, which allows us to derive novel provably optimal sampling strategies, some of which are amenable to closed-form expressions for the estimators.

\section{Conclusion}\label{sec:conclusion}
We studied a fundamental trade-off regarding whether one should
favor a larger number of samples of a single feature (marginal observations) over fewer samples of multiple features (joint observations) in order to effectively estimate   quantities of interest in the presence of resource constraints.  As samples and features play a role in any learning task, and resource costs are an integral part of distributed learning, we believe that the results presented in this work set the ground for a number of interesting directions for future work in the space between networking and machine learning.  
Among those, we envision extending the results to scenarios beyond the bivariate Gaussian distribution, and accounting for classification as opposed to estimation tasks.

\bibliographystyle{IEEEtran}
\bibliography{ref}
\end{document}